\documentclass{aa}
\usepackage{psfig}
\def\h{\hbox{$^{\rm h}$}}
\def\m{\hbox{$^{\rm m}$}}

\def\degr{\hbox{$^\circ$}}
\def\amin{\hbox{$^\prime$}}

\def\fs{\hbox{$.\!\!^{\rm s}$}}

\def\farcm{\hbox{$.\mkern-4mu^\prime$}}
\def\farcs{\hbox{$.\!\!^{\prime\prime}$}}               

\begin{document} 
\thesaurus{06(08.02.2;08.09.2; 08.14.2)}
\title{The eclipsing Cataclysmic Variable GS
Pavonis: Evidence for disk radius changes
\thanks{Based on observations with the 0.9m Dutch telescope at ESO, Chile}}
\author{P.J. Groot\inst{1}
\and T. Augusteijn\inst{2}
\and O. Barziv\inst{1,3}
\and J. van Paradijs\inst{1,4}}
\offprints{Paul Groot (paulgr@astro.uva.nl)}
\institute{Astronomical Institute `Anton Pannekoek'/ CHEAF, Kruislaan
403, 1098 SJ, Amsterdam, The Netherlands
\and European Southern Observatory, Casilla 19001, Santiago 19, Chile
\and European Southern Observatory, Karl-Schwarzschildstr. 2,
D-85748, Garching-bei-M\"unchen, Germany
\and Physics Department, University of Alabama in Huntsville,
Huntsville, USA}
\date{Received date, accepted date}
\authorrunning{P.J. Groot et al.}
\titlerunning{The Eclipsing Cataclysmic Variable GS Pavonis}
\maketitle

\begin{abstract} 
We have obtained differential time series photometry of the cataclysmic
variable GS Pavonis over a timespan of 2 years. These show that
this system is deeply eclipsing ($\sim$2--3.5 mag) with an
orbital period of 3.72 hr. The eclipse depth and out-of-eclipse light
levels are correlated. From this correlation we deduce that the
disk radius is changing and that the eclipses in the low state are
total. The derived distance to GS Pav is
790$\pm$90 pc, with a height above the galactic plane of
420$\pm$60 pc. We classify GS Pav as a novalike system.
\end{abstract}

\section{Introduction}

Eclipsing non-magnetic Cataclysmic Variables (CVs) 
(for a review see Warner 1995, hereafter
W95) are of particular interest not only because the masses of both 
stars can be determined, but
especially because studying their eclipses, e.g., by the eclipse
mapping method (Horne 1985), gives the
opportunity to learn more about the physics of accretion disks. 
In this {\sl Letter} we report that GS Pav is an
eclipsing CV, that shows substantial disk radius changes.  

GS Pav was first discovered by Hoffmeister (1963) who denoted it
as star S7040 and gave the comment `raschwechselnd' (rapidly
varying). It was classified as a dwarf nova type CV in the GCVS and in
the catalogue of Downes, Webbink and Shara (1997), who also give a
finding chart for the object. It was selected for our observations as 
a possible member of the halo population (Augusteijn, 1994).
Zwitter and Munari (1995) show that it has a normal CV spectrum. 
We determined its location at RA= 20\h08\m07\fs58, 
Dec=--69\degr48\amin58\farcs1 (J2000), almost identical to that of
 Downes, Webbink and Shara (1997). 

\section{Observations}
Photometric observations were obtained with the Dutch 0.9m telescope
at ESO La Silla, Chile. A log of the observations is given in 
Table\ \ref{tab:log}. All observations were made with a 512x512 TEK
CCD detector, using a Bessel $V$ filter. Standard
flatfielding and debiasing were applied to all observations. A
photometric calibration was obtained on September 6, 1993 using the 
standard star EG 21 (Landolt 1992). 
Table\ \ref{tab:ref} gives the coordinates and magnitudes of the
reference stars we have used. 

\begin{table}
\caption[]{Log of V-band observations of GS Pav.\label{tab:log}}
\begin{tabular}{llll}
Date & Start UT & Integr. Time (s) &No. Obs.\\[2mm]\hline\\[-2mm]
Sept 5, 1993 & 01\h50\m & 120 & 82 \\
Sept 6, 1993 & 23\h21\m & 120 & 66 \\
Sept 7, 1993 & 23\h11\m & 120 & 48 \\
Sept 8, 1993 & 03\h57\m & 120 & 25 \\
March 22, 1995& 08\h06\m& 240 & 25 \\
March 23, 1995& 07\h15\m& 240 & 30 \\
June 13, 1995& 03\h20\m & 240 & 61 \\
July 11, 1995& 05\h40\m & 240 & 65 \\
July 25, 1995& 04\h23\m & 120 & 24\\
Aug  16, 1995& 04\h52\m & 120 & 54\\
Sept 15, 1995& 03\h23\m & 240 & 21 \\
Sept 19, 1995& 02\h50\m & 240 & 20\\
\hline
\end{tabular}
\end{table}

\begin{table}
\begin{minipage}{8.8cm}
\caption[]{Reference stars used for the differential photometry of GS
Pav. \label{tab:ref}}
\begin{tabular}{l@{\ \ \ }l@{\ \ \ }l@{\ \ \ }l@{\ \ \ }l}
No. & Name & RA (J2000) & Dec (J2000) & V\footnote{The quoted errors
reflect the internal errors in the brightness measurements of the stars
which do not include a 0.1 mag uncertainty in the transformation to
standard magnitudes.}\\[2mm]\hline\\[-2mm]
1	&GAB J200816--6949&20\h08\m16\fs47&--69\degr49\farcm36\farcs0&17.31(4)\\
2	&GAB J200810--6949&20\h08\m10\fs49&--69\degr49\farcm34\farcs0&17.41(4)\\
3	&GAB J200804--6949&20\h08\m04\fs83&--69\degr49\farcm40\farcs4&17.42(4)\\
4	&GAB J200755--6949&20\h07\m55\fs45&--69\degr49\farcm14\farcs4&17.09(4)\\
5	&GAB J200802--6948&20\h08\m02\fs49&--69\degr48\farcm57\farcs6&18.54(7)\\
6	&GAB J200800--6948&20\h08\m00\fs64&--69\degr48\farcm49\farcs5&16.33(3)\\
7	&GAB J200757--6948&20\h07\m57\fs63&--69\degr48\farcm19\farcs6&17.87(5)\\
8	&GAB J200810--6947&20\h08\m10\fs41&--69\degr47\farcm55\farcs4&15.45(2)
\end{tabular}
\end{minipage}
\end{table}

\section{Photometric ephemeris \label{sec:res}}
Arrival times of mid-eclipse were determined by fitting a Gaussian
profile to the eclipses, and by determining the mid-point between the
points of steepest ascent and descent. The final arrival times listed
in Table 3 were taken as the average of the results from these two 
methods. We estimate the accuracy of these arrival times to be 5$\cdot$10$^{-4}$
days, which corresponds to the typical difference between the results 
from the two methods.

A linear fit to the arrival times listed in Table 3 yields the 
following ephemeris:
\begin{equation}
{\rm HJD}_{\rm min} = 244\,9711.17388(17) +
0.155269817(87)\cdot N,
\end{equation} 
with $N$ the cycle number. The error estimates for the parameters are
scaled to give a $\chi^2_{\rm red}$ = 1.0. The rms value of the
arrival times around the fit is 6.4$\cdot$10$^{-4}$ days, 
which is in reasonable agreement with our error estimate.

\begin{table}
\caption[]{Times of arrival, deduced cycle numbers and the observed
minus computed (O--C) residuals for the observations of GS Pav
\label{tab:eclipse}}
\begin{tabular}{llr}
Cycle No. & HJD$_{\rm min}$--244\,0000 & O--C (days)\\[2mm]\hline\\[-2mm]
--3062  &  9235.7370 &  --0.70$\cdot$10$^{-3}$ \\
--3050  &  9237.6014 &    0.46$\cdot$10$^{-3}$ \\
--3044  &  9238.5330 &    0.44$\cdot$10$^{-3}$ \\
--3043  &  9238.6876 &  --0.23$\cdot$10$^{-3}$ \\
565  &  9798.9010 &  --0.32$\cdot$10$^{-3}$ \\
571  &  9799.8340 &    1.05$\cdot$10$^{-3}$ \\
1098  &  9881.6603 &    0.16$\cdot$10$^{-3}$ \\
1099  &  9881.8147 &  --0.71$\cdot$10$^{-3}$ \\
1279  &  9909.7641 &    0.13$\cdot$10$^{-3}$ \\
1280  &  9909.9191 &  --0.14$\cdot$10$^{-3}$ \\
1369  &  9923.7376 &  --0.66$\cdot$10$^{-3}$ \\
1511  &  9945.7857 &  --0.87$\cdot$10$^{-3}$ \\
1704  &  9975.7542 &    0.56$\cdot$10$^{-3}$ \\
1729  &  9979.6362 &    0.82$\cdot$10$^{-3}$ \\
\hline
\end{tabular}
\end{table}

\section{Mass ratio, width of the eclipse and inclination \label{sec:mass}}

If the secondary follows the lower main-sequence 
standard mass-period relation (W95): 
\begin{equation}
M_2 = 0.065\, P_{\rm orb}^{5/4}({\rm h}) \ \ \ \ 1.3\leq\, P_{\rm
orb}({\rm h})\leq 9, 
\end{equation}
with $M_2$ the mass of the secondary, $M_2$ is $\sim 0.34$
M$_{\odot}$. Since the mass ratio, $q$=$M_2$/$M_1$, has to be
smaller than 2/3 for stable mass transfer (W95) 
and the mass of the primary can at most be the Chandrasekhar mass
(1.4 M$_\odot$), the mass ratio is limited to the range 
$0.67 \geq q \geq 0.24$. 

\begin{figure*}
\centerline{\psfig{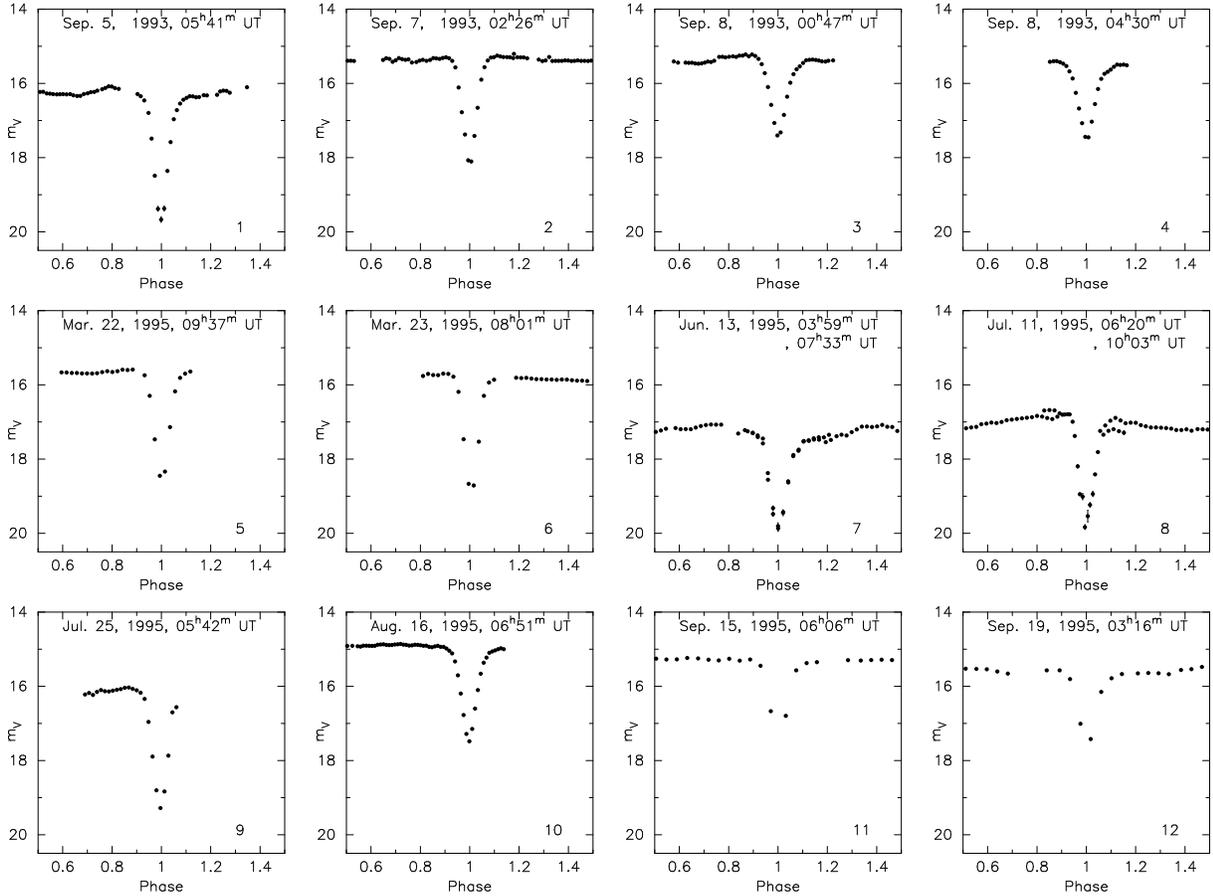}}
\caption[]{The phasefolded V-band eclipse light curves for the 12 epochs. The
heliocentric corrected time of mid-eclipse of the observations is given in UT. \label{fig:eclipse}}
\end{figure*}

Figure\ \ref{fig:eclipse} shows the phase folded eclipse light curves
for the 12 epochs listed in Table\ \ref{tab:log}. The light curves of
June 13 and July 11, 1995 contain two eclipses each. 

The width of the eclipse ($\Delta\varphi$) can be estimated by the phases
of steepest descent and ascent. For GS Pav we measure a mean 
$\Delta\varphi$ = 0.064 $\pm$ 0.005, where the error is the scatter on
the average of all measurements.
From this value and the range in mass ratio's we deduce (Horne 1985) a
range in orbital inclination of 74\degr $<i<$ 83\degr. 

\section{Correlation between eclipse depth and out-of-eclipse
light \label{sec:corr}}

\begin{figure}
\centerline{\psfig{figure=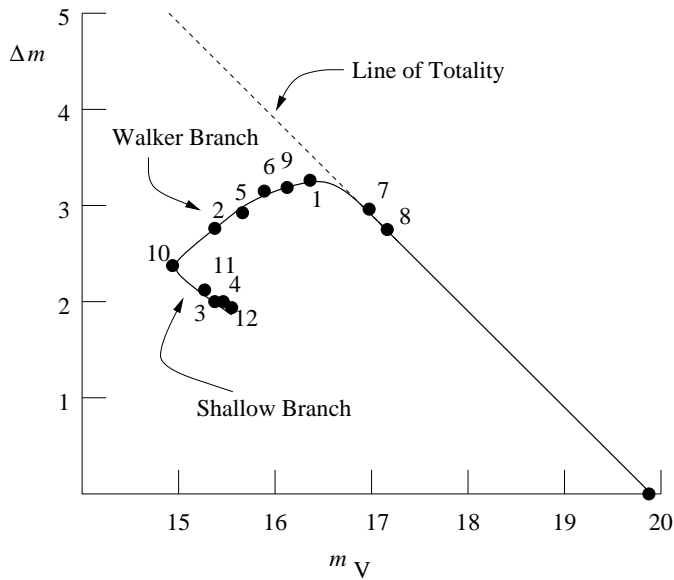,width=8.8cm,angle=-90}}
\caption[]{The depth of the eclipse in GS Pav as function of the
out-of-eclipse light. The numbers correspond to the eclipses
in Fig.\ \ref{fig:eclipse}. 
\label{fig:cycle}}
\end{figure}

Figure\ \ref{fig:eclipse} shows that GS Pav does not have a constant 
out-of-eclipse magnitude. In our observations the
source varied between V$\sim$14.9
and V$\sim$ 17.1, being mostly at the bright end. Also 
the eclipse depth is not constant. We have investigated
if these two variations are correlated. Since the eclipses are well
represented by Gaussian functions, we have taken the zero-level and
depth of the Gaussians, used to determine the
times of mid-eclipse, as estimates of the depth in magnitudes of mid-eclipse
($\Delta m$) and the out-of-eclipse light level ($m_V$) (Fig.\
\ref{fig:cycle}). The numbers in Fig.\ \ref{fig:cycle} refer to the
eclipses as shown in Fig.\ \ref{fig:eclipse}. 

In the following we will make a distinction between the
observed radius of the accretion disk, which is the size we infer from
our observations, and the physical radius of the accretion disk. The
difference between these two is determined by the fractions of the
optically thick and thin parts of the accretion disk and the
brightness distribution across the disk which determines what parts 
are visible in the chosen passband (here in the V-band). A changing
brightness distribution can mimic a change in the physical disk
radius, when observed in only one band. 

In Fig.\ \ref{fig:cycle} the straight line labeled 'Line of Totality' shows
what the correlation between $\Delta m$ and $m_{\rm V}$ looks
like for a system in which the eclipse is total. A total eclipse means
that the observed size of the accretion disk is smaller than the size
of the secondary and that at mid-eclipse the amount of observed light
from the disk is negligible. In a total eclipse the
mid-eclipse light level will be constant and equal to the brightness
of the secondary. Every magnitude of brightening of the disk will
cause a magnitude of deepening of the eclipse: the system
will follow a straight line, with an angle of 45\degr, in the 
$\Delta m$-$m_{\rm V}$ diagram. The position of this line in the diagram will
be different for each individual system, but can be fixed by 
determining the brightness of the secondary. If at any time during our
observation GS Pas was totally eclipsing, then its mid-eclipse light
level will be the brightness of the secondary. 
The minimum level occured on June 13 and July 11, 1995:
$V$=19.9$\pm$0.1. We use this point to
fix the position of the `Line of Totality' in Fig.\ \ref{fig:cycle}. 
We see that points '7' and '8' (which are from June
13 and July 11, 1995) lie on this line. From the fact that the eclipse
depths in point '7' and '8' are different, but the brightness
at mid-eclipse, we conclude that 
at these epochs the eclipse is indeed total.
It follows that the observed minimum of $V$=19.9$\pm$0.1 is the
brightness of the secondary. 
The lack of a flat-bottom in the light curves of point
'7' and '8' shows that the eclipse is only just total, although the
integration time of 4 minutes may be too long to resolve a flat
bottom.  
 
If we now look at the other points in Fig.\ \ref{fig:cycle} we see
that they do not fall on the 'Line of Totality'. 
With increasing out-of-eclipse
light levels, the depth of the eclipse does not increase anymore (as it
would have on the Line of Totality), but decreases. Apparently, between
point '7' and '1' on the track the observed size of the disk increases
to the extent that the eclipse is no longer total, but that part of
the accretion disk remains visible in mid-eclipse. 
With a further increase of the observed accretion disk radius, 
more and more of the disk is visible at mid-eclipse
and the eclipse becomes less and less deep. This behaviour was 
first found by Walker (1963) in his study of RW Tri, which shows
no total eclipses, but does move back and forth on this part of the
track. We have therefore labeled this the `Walker Branch'. 
Another transition, which to our knowledge has never been noted before, 
is the one that happens near point `10' in Fig.\ \ref{fig:cycle}. 
The out-of-eclipse light
level reaches a maximum, after which it declines again (the curve
bends to the right), but the eclipse depth continues to decrease. 
We have labeled this part 
the 'Shallow Branch' because of its decreasing eclipse depth. 

If the change in the observed radius is caused by a change in the physical
size of the disk (rather than a change in the brightness
distribution), then the manifestation of the Shallow Branch may be
explained by the effect of self-eclipses. 
Because the height of the disk is correlated with its physical
size (Frank, King and Raine 1985), self-eclipses of the hot, and
therefore luminous inner parts of the
concave disk by its outer parts, will occur when the physical size of
the disk exceeds a critical value and therefore a critical height.
The out-of-eclipse light level decreases because the more luminous
parts of the accretion disk are self-eclipsed, and at the same time
the eclipse depth can continue to decrease if the disk radius
continues to increase.  To eclipse the inner parts the disk
flaring angle must be \mbox{(90-$i$)\degr} or higher. 
In our case this would mean a flaring angle of 7\degr--16\degr, 
similar to what has been found in other CVs 
(e.g. Robinson et al, 1995). 

The variations appear to trace out a unique track over a substantial
period of time. Changes from one part of the track to another
can occur quite rapidly, e.g. the transition from point '1' to `4'
took place within 4 days. We are currently 
modelling these changes in detail to constrain the geometry of the
system and the changing disk (Groot et al., 1999).

\section{Distance to the system \label{sec:dist}}

With the orbital period of 3.72 hr we can use the $M_{\rm V}-P_{\rm
orb}$ relation of W95 to derive a $M_{\rm V}$ = 10.4 with an estimated
error of 0.2 magnitudes for the secondary (from Fig. 2.46 in W95).
Combined with the apparent magnitude of $V$=19.9$\pm$0.1, this gives a 
distance to GS Pav of 790$\pm$90pc. Given 
its position on the sky, the implied height above the galactic plane
is 420$\pm$60 pc. 
From the distance we can deduce the absolute magnitude of
the system out of eclipse, which is dominated by the accretion disk.
To obtain an absolute magnitude of the accretion disk, we correct 
for the inclination of the disk, 
according to Eq. 2.63 of W95. This correction varies 
from 1.0 mag (for $i$=74\degr) to 2.1 mag (for $i$=83\degr). 
The absolute magnitude ($M_{\rm V}$) lies therefore 
between 6.6 $\geq M_{\rm V} \geq$ 4.4
(for $i$=74\degr) and 5.5 $\geq M_{\rm V} \geq 3.3$ (for i=83\degr). 
Comparison with the mean absolute magnitudes for NLs and 
DNe (Fig. 4.16 and 3.9 from W95) shows that the derived range is in 
the normal regime for NLs, but too bright for DN in quiescence. 

\section{Classification as a novalike system \label{sec:NL}}

From the shape of the light curve and its absolute magnitude, 
we conclude that GS Pav is a 
novalike system, and considering its 
emission line spectrum (Zwitter and Munari, 1995), that it is of the RW Tri
subclass, which is defined as having emission line spectra (W95).
According to the definition in W95 of the VY Scl subclass, as NL systems
having low states in their long term light curves, and showing no DN
outbursts during these low states, we should also classify it as a VY
Scl star. This is supported by its orbital period, since almost all 
known VY Scl stars have periods between 3 and 4 hrs. However, the
physical interpretation as outlined in W95 may not apply to GS Pav. 
In this description a VY Scl system in its low
state has a mass-transfer rate $\dot{M}$, that is lower than the
critical rate, $\dot{M}_{\rm crit}$, below which DN outbursts are
expected to occur, but it does not show these outbursts. This
distinguishes VY Scl systems from Z Cam systems, that do show these
outbursts in their low state. In our observations GS Pav at all times 
seems to have an absolute magnitude which was brighter or equal than
that of Z Cam systems during standstill, where $\dot{M}$ is thought to
be larger than $\dot{M}_{\rm crit}$. We therefore cannot conclude if GS
Pav is a VY Scl system or not. However, it could be that all NL
systems with periods between 3 and 4 hours turn out to have low
and high states if sufficiently long observed (W95).    

An interesting example of a system that may be analogous to GS Pav 
is VZ Scl for which O'Donoghue, Fairall and Warner (1987) 
concluded that the size of the accretion disk has changed from one 
observation to the other. Unfortunately they only observed two
eclipses. Since this system, in the low state, is also totally
eclipsing, a comparison with GS Pav would be of interest. 

\section{Conclusions \label{sec:conc}}

We have shown that GS Pav is a deeply eclipsing cataclysmic variable
with a 3.72 hr period. Based on its photometric behaviour, orbital
period and absolute magnitude we classify the system as a novalike
variable. The depth of the eclipse and the
brightness of the system out-of-eclipse are correlated. 
From this relation we infer that the disk radius changes and that 
in two of our observations the eclipse is total. 
Therefore, the apparent visual magnitude of the
secondary is \mbox{$V$=19.9$\pm$0.1}, giving a distance to the
system of 790$\pm$90 pc.

\end{document}